# The Nature of the Galactic Dark Matter

N.W. Evans

*Theoretical Physics, Department of Physics, 1 Keble Road, Oxford, OX1 3NP*

**ABSTRACT**

Disk-halo models of the Galaxy and LMC are constructed and used to analyse the microlensing data-set. Deflectors in the LMC bar, disk and halo provide an optical depth to microlensing of $\sim 2 \times 10^{-7}$. Deflectors in the Galactic disk and halo contribute $\sim 5 \times 10^{-7}$. The extent, flattening and velocity anisotropy of the dark objects in the Galactic halo are unknown. So, it is crucial to analyse the microlensing data-set with families of models that span the viable ranges of these structural parameters. Also uncertain is the contribution of the Galactic disk to the local circular speed, which affects the normalisation and size of the Galactic halo. Despite all the unknowns, a robust conclusion is that the Galactic and LMC haloes cannot be primarily built from objects in the mass range $10^{-7}\,M_\odot$ - $0.1\,M_\odot$. By contrast, calculations of the baryon mass fraction of the Galactic and LMC haloes are very sensitive to details of the adopted models. This parameter is not constrained by the existing data-set. In particular, it is still possible for the halo to be entirely baryonic and composed of high mass compact objects, like $10^6\,M_\odot$ clusters or black holes.

## 1 INTRODUCTION

The physics of microlensing is the physics of twinkling. The stars twinkle because turbulence and mixing in the upper atmosphere cause patches of high and low refractive index. Equally, stellar images towards the Large Magellanic Cloud (LMC) are microlensed because unseen masses in the dark haloes of our Galaxy and the LMC cause variations in the refractive index (Petrou 1981; Paczyński 1986; Griest 1991). The labours of the two experimental groups monitoring millions of stars in the LMC have already borne fruit with detections of microlensing (Alcock et al. 1993, 1995a; Aubourg et al. 1993, 1995). In this Letter, we show that – irrespective of variations in the flattening, extent and velocity anisotropy of the Galactic halo – the dominant constituent of dark haloes cannot lie in the mass range $10^{-7}\,M_\odot$ - $0.1\,M_\odot$. The Galactic and LMC haloes are not built mainly from brown dwarfs, low mass stars or Jupiters. This result is *robust*, even in the face of the uncertainties as to the contribution of the Galactic halo to the local circular speed, which determines its overall normalisation.

## 2 MODELS OF THE GALAXY AND THE LMC

A crude facsimile of the LMC is provided by embedding an inclined disk in a spherical dark halo (de Vaucouleurs & Freeman 1973; Gould 1993). Westerlund (1990) reckons that the tilt of the LMC disk is $\sim 45°$. The position angle of the line of nodes is more securely known as $\sim 170°$. Let us represent the LMC by a thick exponential disk of scale-length $\sim 1.6$ kpc and scale-height $\sim 0.3$ kpc (Bessell, Freeman & Wood 1986) – that is, by

$$\rho \sim 2.8 \times 10^8 \exp[-0.6(R^2 + 28z^2)^{1/2}]\,M_\odot\,\text{kpc}^{-3}. \qquad (2.1)$$

The LMC centre lies in the direction $(\ell = 280°, b = -33°)$ and at a heliocentric distance of 50 kpc (Westerlund 1990). So, $(R, z)$ are cylindrical polar coordinates with origin at the LMC center and oriented to reproduce the tilt and the line of nodes of the LMC disk. The LMC rotation curve is flattish out to at least $8°$ radius, and possibly even out to $15°$ (Schommer et al. 1992). So, the LMC is probably the possessor of a dark halo that extends out to $\sim 10$ kpc. This halo is taken as a cored and truncated isothermal sphere (Evans 1993)

$$\rho \sim 4.6 \times 10^7\,\frac{3R_c^2 + r^2}{(R_c^2 + r^2)^2}\,M_\odot\,\text{kpc}^{-3}, \qquad (2.2)$$

whose parameters are chosen to make the combined rotation curve of disk and halo flattish and $\sim 80$ km s$^{-1}$ in the outer reaches (Schommer et al. 1992). The core radius $R_c$ is picked as 0.5 kpc and the model is truncated sharply at 10 kpc. The brightest part of the LMC is an off-centered bar (de Vaucouleurs & Freeman 1973) that perhaps contains $\sim 10\%$ of the total mass. Sahu (1994) has argued that this may be an important source of microlenses. It is mimicked by a prolate thick spheroid of total mass $\sim 1 \times 10^9\,M_\odot$. Of course, all the structural parameters of the LMC are uncertain – but the crude model just devised is certainly an improvement on the point mass used by almost all earlier investigators. The microlensing rate depends on the velocity distributions of the sources and the deflectors. The source population is assumed to lie in the disk and bar of the LMC and to be cold and move with roughly the circular speed. The velocity distribution of the halo model (2.2) is provided by Evans (1993). We also need to estimate the proper motion of the LMC centre. In a Galactic coordinate system (axes positive towards the Galactic centre, in the direction of the Galac-of Jones, Klemola & Lin (1994) suggest that the heliocentric motion of the LMC centre is $\sim (90, -360, 110)$ km s$^{-1}$. This result is obtained by repeating their calculations for our different assumed value of the tilt of the LMC disk.

Of course, the heftiest source of deflectors is the dark halo of the Galaxy. Almost everything about the halo is unknown – its flattening, its extent and its velocity anisotropy. So, the data must be analysed within the framework of an ensemble of models, each of which is a possible representation of the Galaxy. A flexible set of halo models with simple velocity distributions is the power-law models (Evans 1994). The halo mass density is

$$\rho = \frac{\overline{A}}{4\pi G q^2} \frac{R^2(1-\beta q^2) + z^2(2-(1+\beta)q^{-2})}{(R^2 + z^2 q^{-2})^{(\beta+4)/2}}. \quad (2.3)$$

This falls off like distance $^{-2-\beta}$. The parameter $q$ controls the shape of the halo. This can be spherical ($q = 1$) or oblate ($q < 1$). The rotation curve can rise ($\beta < 0$), fall ($\beta > 0$) or be flat ($\beta = 0$) at large radii. The parameter $\overline{A}$ determines the depth of the potential well of the halo. As realised by Evans & Jijina (1994), these models allow the construction of an assemblage of simple disk-halo representations of the Galaxy (see also Alcock et al. 1995b; Kan-ya, Nishi & Nakamura 1995). We shall investigate the microlensing properties of three families of models. These are:

(1) A sequence of spherical ($q = 1$), untruncated halo models in which the velocity distribution changes from tangential anisotropy ($\gamma = 10$) through isotropy ($\gamma = 0$) to radial anistropy ($\gamma = -1$). The dependence of the velocity distribution on the anisotropy parameter $\gamma$ is given in eqs (5.4) and (5.6) of Evans (1994).

(2) A sequence of flattened, untruncated halo models in which the ellipticity changes from round or E0 to highly flattened or E6. The simplicity of spherical models has caused them to be widely used (e.g., Paczyński 1986; Griest 1991). But, N-body simulations of gravitational collapse (e.g., Dubinski & Carlberg 1991) indicate that dark halos may be typically flattened.

(3) A sequence of spherical ($q = 1$), truncated halo models in which the extent changes from 30 kpc to 50 kpc. Almost certainly, the Galactic halo extends out to 30 kpc. Evidence from the kinematics of Galactic satellites suggests that the halo may extend beyond 50 kpc, encompassing the LMC (e.g., Fich & Tremaine 1991).

The microlensing data-set has already been analysed with spherical models (e.g., Alcock et al. 1995a; Aubourg et al. 1995). While this Letter was in preparation, Alcock et al. (1995c) provided an interesting analysis of the microlensing data-set with flattened Galactic halo models – but they did not consider the effects of truncation and velocity anisotropy, nor the contribution of the LMC halo.

In each case, the Galactic disk is taken as

$$\rho \sim 6.2 \times 10^8 \exp[-0.28(R^2 + 144z^2)^{1/2}] \, M_\odot \, \text{kpc}^{-3}. \quad (2.4)$$

This disk has an exponential scale-length of 3.5 kpc, while the axis ratio is chosen as 1/12 (Gilmore, King & van der Kruit 1989). An important source of uncertainty is the contribution of the disk to the local circular speed. This affects the overall normalisation of the dark halo and hence the number of events expected towards the LMC. Here, the value for the local column density of the Galactic disk at the Sun is the "canonical" one of $\sim 71 M_\odot$ pc$^{-2}$. This is in agreement with studies of the vertical kinematics of tracer populations (Kuijken & Gilmore 1991). However, the microlensing data toward the Galactic bulge may suggest that such a value is misleadingly low (Griest et al. 1995). So, we outline how our results change if the Galactic disk is "maximal" and the local column density is $\sim 100\, M_\odot$ pc$^{-2}$. In each case, the disk is matched with a power-law halo such that the combined rotation curve is flattish and $\sim 220$ km s$^{-1}$ outwards from the solar circle. The velocity distribution of deflectors in the Galactic halo (2.3) is taken from Evans (1994). The deflectors in the disk are assumed to move in cold, circular orbits.

**Figure 1.** The characteristic masses of halo objects excluded by the experimental results of Alcock et al. (1993, 1995a) are shown as shaded regions. The logarithm of the mass (in units of the solar mass) is plotted horizontally. The vertical axes show the effects of varying the unkown structural parameters of the halo. The velocity distribution of the dark objects may be tangentially anisotropic ($\gamma = 10$) or radially anisotropic ($\gamma = -1$). The halo may be as round as E0 or as flat as E6. It may extend to 30 kpc or beyond. The Galactic disk is canonical. Mass ranges excluded in all three diagrams can be discarded at the 95% confidence level, irrespective of our ignorance of the details.

## 3 ANALYSIS OF THE MICROLENSING DATA-SET

The optical depth to microlensing towards the LMC has contributions from deflectors in the Galactic disk and halo, as well as contributions from the disk, bar and halo of the LMC. The former depends on the extent and flattening of the Galactic halo, but is typically $\sim 5 \times 10^{-7}$ (c.f., Griest 1991). The latter contributes an optical depth of $\sim 2 \times 10^{-7}$. So, the LMC makes a substantial contribution ($\sim 30\%$) to the total optical depth to microlensing. Its effect must be taken into account in data analysis and modelling. Gould (1993) has ingeniously suggested that the signature of the LMC (as opposed to Galactic) microlenses is a variation in the optical depth across the face of the LMC disk. This happens because lines of sight to sources in the far, west side of the LMC disk are longer than those to sources in the near, east side. At least in our model, this effect is unmeasurable. Partly, it is masked by contributions from dark objects in both the Galactic halo and the LMC disk and bar. Partly, our LMC disk is more face–on than assumed by Gould (1993), as we have preferred to use Westerlund's (1990) more recent value of the LMC tilt (45°) rather than the older value of 27° (de Vaucouleurs & Freeman 1973).

**Figure 2.** As Figure 1, but now the Galactic disk is maximal.

Let us assume that the Galactic and LMC haloes are completely built from baryonic dark objects with a characteristic mass $M$ (in units of the solar mass). The differential rate with respect to timescale $t_0$ is computed using the formalism of Kiraga & Paczyński (1994). The contributions of all deflectors in the disk and halo of the Galaxy, as well as the disk, bar and halo of the LMC are incorporated. At every timescale, the differential rate is multiplied by the efficiency of the microlensing experiment at that timescale. The efficiency curves are available in Alcock et al. 1995a. A final quadrature over all timescales gives the rate as observed by the experimentalists. This enables us to calculate what the experiments should have detected in our theoretical models.

Alcock et al. (1993, 1995a) monitored 8.6 million stars in the LMC over a period of 1.1 years and discovered 3 microlensing events. Models predicting in excess of 7.7 events are excluded at the 95% confidence level. Such forbidden models are shown as shaded regions in Figure 1. The logarithm of the characteristic mass $M$ of the dark objects is plotted horizontally. The vertical axes in Figure 1 show the possible ranges of the unknown quantities – velocity anisotropy, flattening and halo extent. Despite all the uncertainties, the dominant constituent of the Galactic and LMC haloes cannot lie in the mass range $10^{-3.75} M_\odot$ - $10^{-0.5} M_\odot$ because this region is excluded in all three diagrams. Suppose, however, the Galactic disk is maximal and the local column density is $\sim 100 M_\odot \, pc^{-2}$. Figure 2 shows the results of repeating the calculations with a maximal Galactic disk and hence smaller halo. The excluded region shrinks slightly – the mass range $10^{-3.75} M_\odot$ - $0.1 M_\odot$ is now ruled out at the 95% confidence level.

**Figure 3.** As Figure 1, but for the experimental results of Aubourg et al. (1993, 1995).

Aubourg et al. (1995) have analysed the light-curves of 82 thousand stars over a period of 10 months with up to 46 measurements per night. They searched for short timescale microlensing events. None were found. This null result can be contrasted directly with the predictions of our theoretical models using the efficiency curves reported by Queinnec (1994). Models yielding in excess of 3.0 events can be discarded at the 95% confidence level. Such rejected models are shown in Figure 3 as shaded regions. Again, by looking at the region excluded in all three diagrams, we see that the major contributor to the Galactic and LMC haloes cannot have a characteristic mass in the range $10^{-7} M_\odot$ - $10^{-3.75} M_\odot$. By combining the results of both the experiments, we conclude that – *irrespective of the uncertainties in the Galactic models* – dark haloes are not predominantly built from objects with masses in the range $10^{-7} M_\odot$ - $0.1 M_\odot$.

So far, only the data on the rate have been exploited. The additional information on the timescale of the three events observed by Alcock et al. (1995a) allows estimation of the mass fraction of the halo that is baryonic (see Alcock et al. 1995c). Unfortunately, a very wide range of values for the baryon mass fraction is still possible. Partly, this is because of Poisson noise in the sparse data-set. Partly, this is because the maximum likelihood calculations are very sensitive to uncertain details of the Galactic models – particularly whether the Galactic disk is canonical or maximal. Partly, this is because the microlensing searches are not responsive to very massive baryonic objects – such as $10^6 \, M_\odot$ black holes (Lacey & Ostriker 1985) or dark clusters (Carr & Lacey 1987; Carr 1994). The timescales of events produced by such lenses are inordinately long, typically $10^2$ or $10^3$ years. It is still possible that the halo is entirely baryonic and dominated by such very massive objects. So, no robust conclusions can presently be drawn from the data-set about the baryon mass fraction.

## 4 CONCLUSIONS

The disk-halo models of the Galaxy and the LMC described in this Letter provide a sophisticated framework to analyse the microlensing data-set. It is important to realise that the LMC halo is a source of lenses second in significance only to the Galactic halo itself. Deflectors in the LMC halo, disk and bar provide $\sim 30\%$ of the total optical depth to microlensing. The effect of the LMC has to be taken into account in the data analysis and modelling.

The flattening, extent and velocity anisotropy of the Galactic halo are all unknown. The contribution of the Galactic disk to the circular speed is also uncertain. It is crucial to analyse the microlensing data-set with a realistic family of models that span the possibilities and to search for robust conclusions. The main result of this paper is that – for all reasonable choices of models – the Galactic and LMC haloes cannot be built from objects in the range $10^{-7} \, M_\odot$ - $0.1 \, M_\odot$. A secure conclusion from the existing data-set is that haloes composed primarily of brown dwarfs, low mass stars or Jupiters are completely ruled out. It is not possible to make any firm estimate of the mass fraction of the Galactic and LMC haloes that is baryonic. In fact, both haloes can still be completely baryonic and built from very massive dark objects, for example, black holes or dark clusters (see e.g., Carr 1985).

The effect of the flattening of the halo on the analysis of the microlensing data-set has recently been explored by Alcock et al. (1995c). In our analysis, this seems to be the least important of the uncertainties. By contrast, the effect on the microlensing rate of different velocity distributions has not been explored before. Our calculations suggest that this is one of the most important of the uncertainties. Given a Galactic halo model, the rate can vary by a factor of 2 or more simply by changing the velocity distribution of the deflectors from tangentially to radially anisotropic. This is unfortunate, as the velocity anisotropy of the dark matter is unknown and possibly unknowable.

## ACKNOWLEDGMENTS

I wish to thank Will Sutherland and Kim Griest for helpful conversations and Jim Rich for sending me a copy of F. Queinnec's Ph. D. thesis. I gratefully acknowledge help from Jim Collett with the diagrams.